# Cauchy's invariants revisited. An alternative variational proof

Gerassimos A. Athanassoulis[a,*], Anastasia Sachinidou[a]

[a] *School of Naval Architecture & Marine Engineering, National Technical University of Athens, 9 Iroon Polytechniou st., Zografos, 15780, Greece*

---

**Abstract**

The purpose of this paper is two-fold. First, to provide a straightforward proof of the Cauchy's invariants (CIs) from the particle relabeling symmetry of the action functional for rotational barotropic flows, using pure geometric relabeling. Second, to emphasize the central role of CIs in the study of rotational flows. The first goal is achieved by introducing a new variational approach for deriving invariants from symmetries, which, when applied to the Lagrangian action of rotational flows, leads to a simple and natural proof of CIs. The second goal is clarified by showing that all important conservation theorems of rotational flows, either local or integral, are simple consequences of CIs.

---

## 1. Introduction

In fluid dynamics, invariants are quantities, either fields or functionals, that remain constant during the motion of the fluid. Field invariants are also called *local invariants*, while invariant functionals are usually called *integral invariants*, since they generally have the form of integrals over the flow fields. Invariants (also called conservation laws) reveal important features of the underlying dynamical systems, and are extremely important, especially when closed form solutions cannot be found. There are various methods to obtain invariants. In the dawn of fluid mechanics, some invariants of the flow had been found by direct elaboration of the governing equations. It was this method that Cauchy used to construct the local invariants that bear his name [1, 2]. Cauchy's Invariants (CIs), although very important in the study of rotational flows, have fallen into oblivion for more than a century [1]. Later, after the integration of the works of Sophus Lie and Emmy Noether in the body of mathematical physics (which happens in 1960s, see [3–5]), two systematic methods were developed for finding invariants. Both are based on the reduction of the problem to the construction of point transformations that leave invariant the equations of motion (Lie approach) or the action functional (Noether's approach). Such transformations are called *symmetries* of the problem under study. Of course, finding all the symmetries of a dynamical system is a cumbersome process. In the field of fluid mechanics, several works have been published in that direction. Olver has applied the Lie approach to construct integral invariants of Euler equations [6], as well as the full set of invariants for water waves in the case of irrotational flow [7]. The same approach has been recently applied to the calculation of local invariants of a fourth-order Boussinesq equation [8]. Noether's approach seems to be especially popular in rotational flows [9–20], although most of these works are limited in scope, as they focus only on one or a few conservation laws, or introduce specific, sometimes unnecessary, restrictions. It is worth noticing that the interest of the scientific community in CIs has been revived in the last years [21–23]. Some far-reaching extensions have been recently presented in [24], using advanced techniques based on exterior calculus and geometrical mechanics.

*Corresponding author
*Email address:* makathan@gmail.com (Gerassimos A. Athanassoulis )

In the present paper we prove the classical CIs using only standard variational arguments. More specifically, we develop an alternative variational approach to construct invariants from symmetries, which is related to Noether's approach but, at the same time, it is succinctly different from it as regards the flow of calculations needed to obtain the invariants. We call this approach the *weak formulation approach*, for reasons that become clear in Sections 3 and 4. The symmetry we investigate is intuitively obvious and very easy to prove; it is the *relabeling symmetry* in the Lagrangian description of rotational fluid flows. In this description, the motion of each fluid parcel is modeled as a field $\boldsymbol{x}_{\boldsymbol{a}}(t) = \boldsymbol{x}(\boldsymbol{a}, t)$, where $\boldsymbol{x} = \boldsymbol{x}_{\boldsymbol{a}}(t)$ is the position at time $t$ of the fluid parcel which is identified by the label $\boldsymbol{a}$. It is clear that, changing the labels does not affect the physics of the fluid flow, exactly as changing the names of indices does not affect the dynamics of a discrete mechanical system. This intuitive premise is easily substantiated by direct verification of the invariance of the action functional under relabeling. Using this new variational method in conjunction with the relabeling symmetry, we present a new, clear and straightforward derivation of CIs, and illuminate their central role as a tool for studying ideal rotational flows by deriving from them many important vorticity theorems. Note that, CIs constitute a complete set of local invariants, according to a theorem of Ertel [25].

Relabeling first appeared in the literature under the more restrictive, and somewhat obscure, concept of "exchange of equivalent particles" [9]. Similar points of view were later adopted by some other authors [19, 26], although in most works relabeling is interpreted as a redefinition of labels [11–18, 20] rendering the concept crisp and clear in understanding. Labels are usually assigned a particular meaning for convenience; they are taken to be either the fluid parcels' initial positions [9, 15, 16, 26] or mass labels [11–13, 18–20, 27, 28]. Most authors consider specific types of relabeling, aiming at the derivation of specific, mainly integral, invariants, while to obtain local invariants, as CIs are, the relabeling must have the form of an arbitrary field. The latter approach is adopted in this work, and it is implemented by using the weak formulation approach, which is established in Section 4.

## 2. Formulation of the rotational free-surface flow problem

In the Lagrangian description, the fluid is modeled as an indexed set of fluid parcels, with continuously distributed index. Naturally, this index, also called label, is taken to be vectorial, and is denoted by $\boldsymbol{a} = (a_1, a_2, a_3)$. There is no particular reason to assign any concrete physical meaning to them [18] (p. 56), [20] (p. 3); what is essential is the selected labels to uniquely identify the fluid parcels throughout the whole fluid domain. In this setup, $\boldsymbol{a}$ is an independent variable which belongs to the label space $V^{(\boldsymbol{a})}$. Denoting the physical space occupied by the fluid as $V^{(\boldsymbol{x})}$, the labeling map (point transformation)

$$V^{(\boldsymbol{a})} \ni \boldsymbol{a} \to \boldsymbol{x}(\boldsymbol{a}, t) \in V^{(\boldsymbol{x})} = V^{(\boldsymbol{x})}(t), \tag{2.1}$$

is assumed to be a $C^2$ bi-diffeomorphism. Herein we consider rotational flows of an inviscid barotropic fluid under the action of external conservative forces. The fields involved are the parcels' position $\boldsymbol{x} = \boldsymbol{x}(\boldsymbol{a}, t)$, the density $\rho = \rho(\boldsymbol{a}, t)$ and the pressure $p = p(\boldsymbol{a}, t)$. To formulate the governing equations, use is made of various geometric characteristics of the point transformation (2.1). The most important entity here is the Jacobian matrix of transformation (2.1), denoted by [J] (see Appendix). Extensive use is made of the Jacobian determinant of transformation (2.1),

$$J(\boldsymbol{a}, t) = J = \det[J] = \det\left(\frac{\partial(x_1, x_2, x_3)}{\partial(a_1, a_2, a_3)}\right), \tag{2.2}$$

and the matrix of the cofactors, denoted by $[J_{ij}]$, which is defined by the equation

$$[J_{ij}](\boldsymbol{a}, t) = [J_{ij}] = \begin{pmatrix} J_{11} & J_{12} & J_{13} \\ J_{21} & J_{22} & J_{23} \\ J_{31} & J_{32} & J_{33} \end{pmatrix} = J\big([J]^{-1}\big)^{\mathrm{T}}. \tag{2.3}$$

$J_{ij}$ are the cofactors of the elements $\partial x_i/\partial a_j$ of the Jacobian matrix. The inverse of the Jacobian matrix $[J]$ is denoted by $[J]^{-1}$, while the superscript T stands for the transpose matrix.

The governing differential equations of the bulk flow are the following [18] (Ch. 3), [29] (Sec. 13-14)

$$\textit{Conservation of mass}: \ \rho(\boldsymbol{a}, t)J(\boldsymbol{a}, t) - \rho_0(\boldsymbol{a})J(\boldsymbol{a}, t_0) = 0. \tag{2.4}$$

$$\textit{Conservation of momentum}:$$

$$\rho_0(\boldsymbol{a})J(\boldsymbol{a}, t_0)\underbrace{[\ddot{\boldsymbol{x}}(\boldsymbol{a}, t) + \nabla_{\boldsymbol{x}}\mathrm{P}(\boldsymbol{x}(\boldsymbol{a}, t))]}_{3\times 1} + \underbrace{[J_{ij}]}_{3\times 3}\underbrace{\nabla_{\boldsymbol{a}}\, p(\boldsymbol{a}, t)}_{3\times 1} = 0, \tag{2.5}$$

where $\rho_0(\boldsymbol{a}) = \rho(\boldsymbol{a}, t_0)$, $\nabla_{\boldsymbol{x}} = (\partial/\partial x_1, \partial/\partial x_2, \partial/\partial x_3)^{\mathrm{T}}$, $\nabla_{\boldsymbol{a}} = (\partial/\partial a_1, \partial/\partial a_2, \partial/\partial a_3)^{\mathrm{T}}$ and $\mathrm{P} = \mathrm{P}\,(\boldsymbol{x}(\boldsymbol{a}, t))$ is the potential of any external force field (per unit mass). The gravitational field, corresponding to $\mathrm{P} = gx_3$, is always included in our considerations. In the Lagrangian description of the flow, time differentiation is denoted either by an overdot or by $\partial/\partial t$, which always has the meaning $\partial/\partial t = (\partial/\partial t)_{\boldsymbol{a}=\mathrm{const.}}$.

Since the fluid is assumed to be barotropic, pressure and density are related via the constitutive relation [30] (Sec. 16), [31] (Chapter 2)

$$p = p(\rho) = \rho^2 \partial E(\rho)/\partial\rho, \tag{2.6}$$

where $E = E(\rho)$ is the internal energy per unit mass of the fluid parcel.

Concerning the boundary conditions, we distinguish two types of boundaries; the free-surface $\partial V_{\mathrm{f}}^{(\boldsymbol{a})}$, and the rigid wall $\partial V_{r}^{(\boldsymbol{a})}$. Therefore, we have the following kinematic and dynamic conditions,

$$\dot{\boldsymbol{x}}(\boldsymbol{a}, t)^{\mathrm{T}} \boldsymbol{n}^{(\boldsymbol{a})} = 0 \ on \ \partial V_{r}^{(\boldsymbol{a})}, \tag{2.7a}$$

$$p(\boldsymbol{a}, t) = \bar{p}(\boldsymbol{a}, t) \ on \ \partial V_{\mathrm{f}}^{(\boldsymbol{a})}, \tag{2.7b}$$

where $\boldsymbol{n}^{(\boldsymbol{a})}$ is the outward unit normal vector on the corresponding boundaries and $\bar{p}(\boldsymbol{a}, t)$ is a known applied pressure. In the sequel we consider $\bar{p}(\boldsymbol{a}, t) = 0$, unless otherwise stated.

**Remark 1.** In this paper, the non-scalar equations are usually written and treated in matrix form, as e.g. Eq. (2.5). The matrix formalism is especially convenient in the Lagrangian description since the matrices [J] and $[J_{ij}]$ intervene very frequently in the calculations. The vector fields are considered to be vertical $3 \times 1$ matrices. ■

The variational formulation of the inviscid fluid flow problem in the Lagrangian description (either with a free surface or without) can be easily constructed, since it is a straightforward extension of that for discrete systems. However, density is continuously distributed in fluids and fluid parcels deform as they move, and thus, conservation of mass (2.4) needs to be taken into account too. Following [14–16, 18, 26, 28], the primitive action functional of the system has a single functional argument, namely the position field $\boldsymbol{x}$, and reads as follows

$$\boldsymbol{S}_P[\boldsymbol{x}(\cdot,\cdot)] = \\ = \int_{t_0}^{t_1} \int_{V^{(\boldsymbol{a})}} \left( \frac{1}{2} \dot{\boldsymbol{x}}^2(\boldsymbol{a}, t) - E(\rho(\boldsymbol{a}, t)) - \mathrm{P}[\boldsymbol{x}(\boldsymbol{a}, \mathbf{t})] \right) \rho(\boldsymbol{a}, t) J(\boldsymbol{a}, t) dV^{(\boldsymbol{a})} dt, \tag{2.8}$$

where $[t_0, t_1]$ is an arbitrary time interval. Note that, the density field $\rho$, appearing in Eq. (2.8), is not an independent functional argument. It is defined in terms of the position field $\boldsymbol{x}$ via conservation of mass, Eq. (2.4), that is,

$$\rho(\boldsymbol{a}, t) \equiv \rho[\boldsymbol{x}(\boldsymbol{a}, t)] = \rho_0(\boldsymbol{a}) J(\boldsymbol{a}, t_0) / J(\boldsymbol{a}, t), \tag{2.9}$$

where the position field $\boldsymbol{x}$ appears in the Jacobian $J$ through the partial spatial derivatives $\partial x_i/\partial a_j$. By Hamilton's principle, the dynamics of the problem under consideration is governed by the variational equation

$$\delta_{\boldsymbol{x}} \boldsymbol{S}_P[\boldsymbol{x}(\cdot,\cdot)] = 0, \tag{2.10}$$

with isochronous variations. Using the standard methods of Calculus of Variations, Eq. (2.10) yields as Euler-Lagrange equation the momentum equation (2.5), along with the boundary conditions (2.7).

## 3. The general variation of a functional and Noether's First Theorem

Consider a generic action functional (not necessarily the functional (2.8)), of the form

$$\mathcal{S}[\boldsymbol{x}(\cdot,\cdot)] = \int_{t_0}^{t_1} \int\limits_{V^{(\boldsymbol{a})}} L\left(\boldsymbol{x}(\boldsymbol{a},\mathbf{t}), \dot{\boldsymbol{x}}(\boldsymbol{a},\mathbf{t}), \nabla_{\boldsymbol{a}}\boldsymbol{x}(\boldsymbol{a},\mathbf{t})\right) dV^{(\boldsymbol{a})} dt, \tag{3.1}$$

which is subject to a continuous infinitesimal transformation (of independent and dependent variables)

$$\tilde{t} = t + \delta t\left(\boldsymbol{a}, t, \boldsymbol{x}, \dot{\boldsymbol{x}}, \nabla_{\boldsymbol{a}}\boldsymbol{x}\right), \tag{3.2a}$$

$$\tilde{\boldsymbol{a}} = \boldsymbol{a} + \delta \boldsymbol{a}\left(\boldsymbol{a}, t, \boldsymbol{x}, \dot{\boldsymbol{x}}, \nabla_{\boldsymbol{a}}\boldsymbol{x}\right), \tag{3.2b}$$

$$\tilde{x}_i(\tilde{\boldsymbol{a}}, t) = x_i(\boldsymbol{a}, t) + \delta x_i\left(\boldsymbol{a}, t, \boldsymbol{x}, \dot{\boldsymbol{x}}, \nabla_{\boldsymbol{a}}\boldsymbol{x}\right),\ i = 1,2,3, \tag{3.2c}$$

with variations $\delta t$, $\delta \boldsymbol{a}$ and $\delta \boldsymbol{x}$ being sufficiently smooth arbitrary functions. The general (or total) variation, under transformation (3.2) is given, in vector notation, by the formula (see, e.g., [32] (Sec. 35), [33] (Ch. 4, Sec. 1), [34] (Sec. 13.7))

$$\begin{aligned}\delta\mathcal{S}[\boldsymbol{x}(\cdot,\cdot)] = & \underbrace{\int_{t_0}^{t_1} \int\limits_{V^{(\boldsymbol{a})}} \left[\frac{\partial L}{\partial \boldsymbol{x}} - \sum_{j=1}^{3} \frac{\partial}{\partial a_j}\frac{\partial L}{\partial_j \boldsymbol{x}} - \frac{\partial}{\partial t}\frac{\partial L}{\partial \dot{\boldsymbol{x}}}\right] \cdot \bar{\delta}\boldsymbol{x} dV^{(\boldsymbol{a})} dt}_{\delta\mathcal{S}_{\mathrm{EL}}[\boldsymbol{x}(\cdot,\cdot)]} + \\ & \underbrace{+ \int_{t_0}^{t_1} \int\limits_{V^{(\boldsymbol{a})}} \left[\frac{\partial}{\partial t}\left(L\delta t + \frac{\partial L}{\partial \dot{\boldsymbol{x}}} \cdot \bar{\delta}\boldsymbol{x}\right) + \sum_{j=1}^{3} \frac{\partial}{\partial a_j}\left(L\delta a_j + \frac{\partial L}{\partial_j \boldsymbol{x}} \cdot \bar{\delta}\boldsymbol{x}\right)\right] dV^{(\boldsymbol{a})} dt,}_{\delta\mathcal{S}_{\mathrm{BD}}[\boldsymbol{x}(\cdot,\cdot)]}\end{aligned} \tag{3.3}$$

which is called the *fundamental variational formula.* The part of the variation denoted by $\delta\mathcal{S}_{\mathrm{EL}}$ contains the Euler-Lagrange expressions multiplied by the local variation $\bar{\delta}\boldsymbol{x}$ (see below for definition), and thus describes the bulk contributions to the total variation. The part of the variation denoted by $\delta\mathcal{S}_{\mathrm{BD}}$ describes the boundary contributions since, applying the Divergence Theorem, it is, in fact, a boundary integral containing terms evaluated at the time endpoints, $t_0$ and $t_1$, and a time integral over the net outward flux through the boundary $\partial V^{(\boldsymbol{a})}$.

In Eq. (3.3), $\bar{\delta}\boldsymbol{x}$ is the variation of the field $\boldsymbol{x}$ taken at a fixed point $(\boldsymbol{a}, t)$, and thus, we refer to it as the *local variation.* As shown in [32–34], the *total variation* $\delta\boldsymbol{x}$, under the general transformation (3.2), is related to the local variation $\bar{\delta}\boldsymbol{x}$ by the formula (in vector notation)

$$\bar{\delta}\boldsymbol{x} = \delta\boldsymbol{x} - \dot{\boldsymbol{x}}(\boldsymbol{a}, t)\delta t - (\delta\boldsymbol{a} \cdot \nabla_{\boldsymbol{a}})\boldsymbol{x}(\boldsymbol{a}, t). \tag{3.4}$$

The somewhat complicated formula (3.4) is due to the fact that the total variation of the field $\boldsymbol{x}$ consists of the local variation $\bar{\delta}\boldsymbol{x}$ plus the terms $\dot{\boldsymbol{x}}\delta t$ and $(\delta\boldsymbol{a} \cdot \nabla_{\boldsymbol{a}})\boldsymbol{x}$ stemming from the variations of the time and space boundaries, respectively.

Now if the action functional (3.1) is invariant under the infinitesimal transformation (3.2), its total variation should vanish, $\delta\mathcal{S} = 0$, which is equivalent to (see Eq. (3.3))

$$\delta\mathcal{S}_{\mathrm{EL}}[\boldsymbol{x}(\cdot, \cdot)] + \delta\mathcal{S}_{\mathrm{BD}}[\boldsymbol{x}(\cdot, \cdot)] = 0. \tag{3.5}$$

Eq. (3.5) is known as the Rund-Trautman identity [34] (Sec. 6.2). This consequence of the invariance can be further explored if the fields involved are extremals, that is, if they satisfy the Euler-Lagrange equations of motion. Then, the term $\delta\mathcal{S}_{\mathrm{EL}}$ vanishes identically, and Eq. (3.5) reduces to the equation

$$\delta\mathcal{S}_{\mathrm{BD}}[\boldsymbol{x}(\cdot, \cdot)] = 0. \tag{3.6}$$

Eq. (3.6) is Noether's First Theorem, proved by Noether in 1915 and published in her 1918's article [35]. The integrand of the left-hand side of Eq. (3.6) (see also Eq. (3.3)) is known as Noether's current, and provides a conservation law in field-theoretic considerations [32–34]. Thus, Noether's First Theorem entails that, for every continuous infinitesimal transformation (3.2) which leaves invariant the action functional (3.1), there exists a conservation law which, on the extremals, is encapsulated in Eq. (3.6).

## 4. An alternate variational derivation of conservation laws

An alternative approach to derive conservation laws, which seems to have been unnoticed up to now, is described in the following proposition:

**Proposition 1**: Suppose that the action functional (3.1), is invariant under the family of continuous infinitesimal transformations (3.2), that is, Eqs. (3.2) define a symmetry of the functional. Then, the identity

$$\delta\mathcal{S}_{\mathrm{EL}} = \int_{t_0}^{t_1} \int_{V^{(\boldsymbol{a})}} \left[ \frac{\partial L}{\partial \boldsymbol{x}} - \sum_{j=1}^{3} \frac{\partial}{\partial a_j} \frac{\partial L}{\partial_j \boldsymbol{x}} - \frac{\partial}{\partial t} \frac{\partial L}{\partial \dot{\boldsymbol{x}}} \right] \cdot \bar{\delta}\boldsymbol{x} dV^{(\boldsymbol{a})} dt = 0, \tag{4.1}$$

where the variation $\bar{\delta}\boldsymbol{x}$ is given by Eq. (3.4), leads to invariant quantities (conservation laws) along the extremals of $\mathcal{S}$. ■

*Proof*: If we assume invariance of the functional under the infinitesimal transformations (3.2), then the direction $(\delta t, \delta\boldsymbol{a}, \delta\boldsymbol{x})$ is a symmetry, and Eq. (3.5) holds true. If, in addition, the involved fields satisfy the Euler-Lagrange equations, then $\delta\mathcal{S}_{\mathrm{EL}} = 0 = \delta\mathcal{S}_{\mathrm{BD}}$. Under these circumstances, the following equivalence holds true

$$\left\{ \begin{array}{l} \delta\mathcal{S}_{\mathrm{EL}} = 0 \\ \text{along the symmetry} \\ (\delta t, \delta\boldsymbol{a}, \delta\boldsymbol{x}) \end{array} \right\} \Longleftrightarrow \left\{ \begin{array}{l} \delta\mathcal{S}_{\mathrm{BD}} = 0 \\ \text{along the symmetry} \\ (\delta t, \delta\boldsymbol{a}, \delta\boldsymbol{x}) \end{array} \right\}, \tag{4.2}$$

which completes the proof of the proposition. For obvious reasons, we will refer to the left-hand side approach as the *weak formulation approach*, while the right-hand side is the classical *Noether's Theorem approach.*

It is worthwhile to comment on the content of the above proposition. If the variation $\bar{\delta}\boldsymbol{x}$ is considered arbitrary, then Eq. (4.1) provides us with the Euler-Lagrange equations, that is, the complete information associated with the critical-point condition of the functional. Now, if the variation $\bar{\delta}\boldsymbol{x}$ is taken along the specific direction $(\delta t, \delta\boldsymbol{a}, \delta\boldsymbol{x})$ (see Eqs. (3.4)) in the "extended phase space" $(t, \boldsymbol{a}, \boldsymbol{x})$, it is natural to expect that Eq. (4.1) will provide us with some restricted amount of information associated with the critical-point condition of the functional. This more specific (restricted) information happens to be a conservation law, as far as the direction $(\delta t, \delta\boldsymbol{a}, \delta\boldsymbol{x})$ leaves the functional invariant.

The proposition presented herein establishes an alternate method, apparently equivalent to Noether's First Theorem, for extracting conservation laws variationally. A practically interesting question is *which one of the two equivalent methods is more convenient to apply*? The answer to this question is by no means obvious or unique. The generality in the structure of those two forms, along with the wide variety of physical problems on which the variational theory applies, render it difficult to provide a generic answer. In the next section we shall see how the weak formulation approach leads directly to CIs, by exploiting the particle relabeling symmetry for rotational flows, without invoking mass labels which are usually considered in the Noether's Theorem approach [11–13, 18–20, 27].

## 5. Particle relabeling symmetry and Cauchy's invariants

### *5.1. The particle relabeling symmetry*

As discussed in Sec.2, the mapping (2.1) defines a labeling of fluid parcels. A relabeling is defined by means of another bi-diffeomorphism,

$$V^{(\tilde{\boldsymbol{a}})} \ni \tilde{\boldsymbol{a}} \to \tilde{\boldsymbol{x}}(\tilde{\boldsymbol{a}}, t) \in V^{(\boldsymbol{x})} = V^{(\boldsymbol{x})}(t), \tag{5.1}$$

where $\tilde{\boldsymbol{a}}$ is a different set of labels. The particle relabeling we consider in this paper is analytically described by a time-independent, regular transformation, whose infinitesimal generator is

$$\tilde{t} = t, \tag{5.2a}$$

$$\tilde{\boldsymbol{a}}(\boldsymbol{a}) = \boldsymbol{a} + \delta\boldsymbol{a}(\boldsymbol{a}), \tag{5.2b}$$

$$\tilde{x}_i(\tilde{\boldsymbol{a}}, t) = x_i(\boldsymbol{a}, t), i = 1, 2, 3. \tag{5.2c}$$

Eq. (5.2c) expresses the physical fact that, the components of the position vector $\boldsymbol{x} = \boldsymbol{x}(\boldsymbol{a}, t)$ of the particle designated by $\boldsymbol{a}$, are invariant with respect to particle relabeling [15, 16]. Note that the choice of time-dependent relabeling, with $\delta\boldsymbol{a} = \delta\boldsymbol{a}(\boldsymbol{a}, t)$, is possible too, and has been performed by some authors [11, 12, 14, 16], but leads to unnecessary complicacies (e.g, Eq. (5.2c) does not apply in this case).

It is not difficult to prove that transformation (5.2) leaves invariant the action functional (2.8). First, scalar fields $\psi(\boldsymbol{a}, t) \equiv \psi\left[\boldsymbol{x}(\boldsymbol{a}, t)\right]$ remain invariant, since according to Eq. (5.2c),

$$\tilde{\psi}(\tilde{\boldsymbol{a}}, t) \equiv \psi\left[\tilde{\boldsymbol{x}}(\tilde{\boldsymbol{a}}, t)\right] = \psi(\boldsymbol{x}, t) = \psi\left[\boldsymbol{x}(\boldsymbol{a}, t)\right] \equiv \psi(\boldsymbol{a}, t). \tag{5.3}$$

This argument applies to the density field $\rho[\boldsymbol{x}(\boldsymbol{a},t)]$ and to the scalar potential of the external force field $\mathrm{P}[\boldsymbol{x}(\boldsymbol{a},t)]$. The velocity field $\dot{\boldsymbol{x}}(\boldsymbol{a},t)$ is also invariant under transformation (5.2), since by Eq. (5.2c) it holds that

$$\dot{x}_i(\boldsymbol{a},t) = \dot{\tilde{x}}_i(\tilde{\boldsymbol{a}}(\boldsymbol{a}),t) = \dot{\tilde{x}}_i(\tilde{\boldsymbol{a}},t). \tag{5.4}$$

Further, using Eq. (A.8c)[1], the volume element of the physical space, $dV^{(\boldsymbol{x})}$, is expressed in terms of the volume elements of the label space, $dV^{(\boldsymbol{a})}$ and $dV^{(\tilde{\boldsymbol{a}})}$, by the formulae

$$dV^{(\boldsymbol{x})} = J(\boldsymbol{a},t)dV^{(\boldsymbol{a})} = \tilde{J}(\tilde{\boldsymbol{a}},t)dV^{(\tilde{\boldsymbol{a}})}. \tag{5.5}$$

Combining Eqs. (5.3)-(5.5), it is directly seen that the relabeling transformation (5.2) leaves invariant the functional (2.8), that is,

$$\delta\mathcal{S}_P[\boldsymbol{x}(\cdot,\cdot)] = 0 \;\text{ under transformation (5.2).} \tag{5.6}$$

Therefore, transformation (5.2) is a symmetry of the functional $\mathcal{S}_P$, which is commonly referred to as the particle relabeling symmetry.

It turns out that the class of relabeling symmetries which is the most appropriate for our purposes are those which satisfy the equation $\det\left(\partial(\tilde{\boldsymbol{a}})/\partial(\boldsymbol{a})\right) = 1$. For infinitesimal transformations, this condition is equivalent with $\nabla_{\boldsymbol{a}} \cdot \delta\boldsymbol{a} = 0$, which leads to the representation [11, 12, 19, 20]

$$\delta\boldsymbol{a}(\boldsymbol{a}) = \nabla_{\boldsymbol{a}} \times \delta\boldsymbol{R}(\boldsymbol{a}), \tag{5.7}$$

where $\nabla_{\boldsymbol{a}}\times$ is the curl operator in the labeling space, and $\delta\boldsymbol{R}$ is an arbitrary, sufficiently smooth vector field.

**Remark 2.** Representation (5.7) is not unique; for instance, one could use the representation [36]

$$\delta\boldsymbol{a}(\boldsymbol{a}) = \nabla_{\boldsymbol{a}}\delta R_1(\boldsymbol{a}) \times \nabla_{\boldsymbol{a}}R_2(\boldsymbol{a}),$$

where $\delta R_1, R_2$ are arbitrary scalar functions. This type of representation is mainly encountered in the case of baroclinic flows ($\delta\boldsymbol{a} = \nabla_{\boldsymbol{a}}\delta R_1 \times \nabla_{\boldsymbol{a}} s$). It arises as a consequence of the invariance of entropy, which must be secured in order for the functional to be invariant [15–17, 26]. ■

[1]Equations designated by (A.*), e.g. (A.3) or (A.8c), are given in the Appendix, at the end of the paper.

*5.2. Cauchy's invariants using the weak formulation approach*

The starting point to find the conservation law that corresponds to the relabeling symmetry, using the weak formulation approach (see Proposition 1), is the equation

$$\int_{t_0}^{t_1} \int_{V^{(\boldsymbol{a})}} \left(\rho_0(\boldsymbol{a})J(\boldsymbol{a},t_0)\left(\underline{\ddot{\boldsymbol{x}}^{\mathrm{T}}(\boldsymbol{a},t)} + \underline{\nabla_{\boldsymbol{x}}^{\mathrm{T}}\mathrm{P}}\right) + \underline{\left([J_{ij}]\nabla_{\boldsymbol{a}} p(\boldsymbol{a},t)\right)^{\mathrm{T}}}\right) \bar{\delta}\boldsymbol{x} dV^{(\boldsymbol{a})} dt = 0. \tag{5.8}$$

The integrand of the above equation is the Euler-Lagrange expression of the equations of motion for the rotational flow problem (see Eq. (2.5)), multiplied by the local variations $\bar{\delta}\boldsymbol{x}$. Applying the general formula (3.4) to relabeling transformation (5.2), we obtain

$$\bar{\delta}\boldsymbol{x} = -[J]\delta\boldsymbol{a}. \tag{5.9}$$

Now, we shall transform the integrand in Eq. (5.8) to a form revealing a conservation law for the examined flow problem. In this conjunction, we elaborate on the product of each underlined quantity in Eq. (5.8) and the local variation $\bar{\delta}\boldsymbol{x}$. We start by studying the term $\ddot{\boldsymbol{x}}^{\mathrm{T}}\bar{\delta}\boldsymbol{x}$. Using Eqs. (5.7) and (5.9), and exploiting the associativity and transposition of matrix multiplication, we obtain

$$\ddot{\boldsymbol{x}}^{\mathrm{T}}\bar{\delta}\boldsymbol{x} = -\ddot{\boldsymbol{x}}^{\mathrm{T}}\left([J]\delta\boldsymbol{a}\right) = -\left(\ddot{\boldsymbol{x}}^{\mathrm{T}}[J]\right)\delta\boldsymbol{a} = -\left([J]^{\mathrm{T}}\ddot{\boldsymbol{x}}\right)^{\mathrm{T}}\left(\nabla_{\boldsymbol{a}} \times \delta\boldsymbol{R}\right). \tag{5.10}$$

The rightmost term of the above equation is reformulated by using the vector identity (see e.g., [37] (p. 822)), written in matrix form (see Remark 1 concerning notation),

$$\boldsymbol{w}^{\mathrm{T}}(\nabla \times \boldsymbol{v}) = (\nabla \times \boldsymbol{w})^{\mathrm{T}}\boldsymbol{v} + \nabla^{\mathrm{T}}(\boldsymbol{v} \times \boldsymbol{w}), \tag{5.11}$$

leading to

$$\ddot{\boldsymbol{x}}^{\mathrm{T}}\bar{\delta}\boldsymbol{x} = -\left(\nabla_{\boldsymbol{a}} \times \left(\underline{[J]^{\mathrm{T}}\ddot{\boldsymbol{x}}}\right)\right)^{\mathrm{T}}\delta\boldsymbol{R} - \nabla_{\boldsymbol{a}}^{\mathrm{T}}\left(\delta\boldsymbol{R} \times \left([J]^{\mathrm{T}}\ddot{\boldsymbol{x}}\right)\right). \tag{5.12}$$

Now, by using simple algebraic manipulations and the identities (A.3) and (A.5), the underlined term $[J]^{\mathrm{T}}\ddot{\boldsymbol{x}}$ is transformed as follows

$$\begin{aligned}[J]^{\mathrm{T}}\ddot{\boldsymbol{x}} = \frac{\partial}{\partial t}\left([J]^{\mathrm{T}}\dot{\boldsymbol{x}}\right) - \frac{\partial[J]^{\mathrm{T}}}{\partial t}\dot{\boldsymbol{x}} &= \frac{\partial}{\partial t}\left([J]^{\mathrm{T}}\dot{\boldsymbol{x}}\right) - \left(\nabla_{\boldsymbol{a}}\dot{\boldsymbol{x}}^{\mathrm{T}}\right)\dot{\boldsymbol{x}} = \\ &= \frac{\partial}{\partial t}\left([J]^{\mathrm{T}}\dot{\boldsymbol{x}}\right) - \frac{1}{2}\nabla_{\boldsymbol{a}}\left(\dot{\boldsymbol{x}}^{\mathrm{T}}\dot{\boldsymbol{x}}\right).\end{aligned} \tag{5.13}$$

Inserting Eq. (5.13) in Eq. (5.12), and using the fact that the curl of a gradient is zero, we obtain

$$\ddot{\boldsymbol{x}}^{\mathrm{T}}\bar{\delta}\boldsymbol{x} = -\left(\nabla_{\boldsymbol{a}} \times \frac{\partial}{\partial t}\left([J]^{\mathrm{T}}\dot{\boldsymbol{x}}\right)\right)^{\mathrm{T}} \delta\boldsymbol{R} - \nabla_{\boldsymbol{a}}^{\mathrm{T}}\left(\delta\boldsymbol{R} \times \left([J]^{\mathrm{T}}\ddot{\boldsymbol{x}}\right)\right). \tag{5.14}$$

Working similarly, we obtain

$$\nabla_{\boldsymbol{x}}^{\mathrm{T}}\mathrm{P}\bar{\delta}\boldsymbol{x} = -\nabla_{\boldsymbol{a}}^{\mathrm{T}}\left(\delta\boldsymbol{R} \times \left([J]^{\mathrm{T}}\nabla_{\boldsymbol{x}}\mathrm{P}\right)\right), \tag{5.15}$$

$$\left([J_{ij}]\nabla_{\boldsymbol{a}}p\right)^{\mathrm{T}}\bar{\delta}\boldsymbol{x} = -\nabla_{\boldsymbol{a}}^{\mathrm{T}}\left(\delta\boldsymbol{R} \times \frac{\nabla_{\boldsymbol{a}}p}{\rho}\right). \tag{5.16}$$

Substituting Eqs. (5.14) – (5.16) into Eq. (5.8), we obtain[2]

$$\begin{aligned}\text{Integrand of Eq.(5.8)} &= -\rho_0(\boldsymbol{a})J(\boldsymbol{a},t_0)\left(\nabla_{\boldsymbol{a}} \times \tfrac{\partial}{\partial t}\left([J]^{\mathrm{T}}\dot{\boldsymbol{x}}(\boldsymbol{a},t)\right)\right)^{\mathrm{T}}\delta\boldsymbol{R}(\boldsymbol{a}) - \\ &-\rho_0(\boldsymbol{a})J(\boldsymbol{a},t_0)\nabla_{\boldsymbol{a}}^{\mathrm{T}}\left[\delta\boldsymbol{R}(\boldsymbol{a}) \times \underline{\left([J]^{\mathrm{T}}\ddot{\boldsymbol{x}}(\boldsymbol{a},t) + [J]^{\mathrm{T}}\nabla_{\boldsymbol{x}}\mathrm{P}(\boldsymbol{x}) + \tfrac{\nabla_{\boldsymbol{a}}p(\boldsymbol{a},t)}{\rho(\boldsymbol{a},t)}\right)}\right].\end{aligned} \tag{5.17}$$

Now the underlined expression in the rightmost term of Eq. (5.17) is easily recognized to be the Euler-Lagrange expression of the flow problem, Eq. (2.5), although in a slightly different form. Since, by assumption, we work along the extremals, the Euler-Lagrange expression vanishes, and thus, the underlined term vanishes too. This permits to rewrite Eq. (5.8) in the following simple form:

$$\int_{t_0}^{t_1}\int_{V^{(\boldsymbol{a})}} \rho_0(\boldsymbol{a})J(\boldsymbol{a},t_0)\left(\nabla_{\boldsymbol{a}} \times \frac{\partial}{\partial t}\left([J]^{\mathrm{T}}\dot{\boldsymbol{x}}(\boldsymbol{a},t)\right)\right)^{\mathrm{T}}\delta\boldsymbol{R}(\boldsymbol{a})dV^{(\boldsymbol{a})}dt = 0. \tag{5.18}$$

Since the time interval $[t_0, t_1]$ and the vector field $\delta\boldsymbol{R}$ are arbitrary, and $\rho_0(\boldsymbol{a})J(\boldsymbol{a},t_0) \neq 0$, standard variational arguments imply that Eq. (5.18) is equivalent with

$$\nabla_{\boldsymbol{a}} \times \frac{\partial}{\partial t}\left([J]^{\mathrm{T}}\dot{\boldsymbol{x}}(\boldsymbol{a},t)\right), \quad \boldsymbol{a} \in V^{(\boldsymbol{a})}. \tag{5.19}$$

[2] A form similar to Eq. (5.17) has also been derived by [14–16, 38] using Noether's Second Theorem. It is referred to as a generalized Bianchi identity and constitutes a stronger conservation law of the fluid problem, since it does not require that the Euler-Lagrange equations of motion be satisfied.

Eqs. (5.19) are the Cauchy's invariants written in Hankel's form [1]. As has been established by [1], these equations, although being pivotal in rotational flows (since most of the well-known vorticity theorems and conservation laws can be derived from them), have been almost forgotten until the end of 20th century. Important consequences of CIs are investigated below.

## 6. Derivation of various theorems of vortical flows from Cauchy's invariants

*6.1. Image fields in the label space*

In studying the consequences of CIs, it is particularly convenient to exploit the images of the physical fields in the label space; see e.g., in [39](Appendix A4), and [19] (Sec. 7.16). The images of the velocity and the vorticity fields, in the label space, are defined by the equations

$$\boldsymbol{V}(\boldsymbol{a},t) \stackrel{\text{def}}{=} [J]^{\mathrm{T}}\dot{\boldsymbol{x}}(\boldsymbol{a},t) = [J]^{\mathrm{T}}\boldsymbol{u}(\boldsymbol{x}(\boldsymbol{a},t),t), \tag{6.1a}$$

$$\Omega(\boldsymbol{a},t) \stackrel{\text{def}}{=} \nabla_{\boldsymbol{a}} \times \boldsymbol{V}(\boldsymbol{a},t). \tag{6.1b}$$

$\Omega(\boldsymbol{a},t)$ is also referred to as the *Lagrangian vorticity* [40], or *Beltrami's material vorticity* [41]. Note that, by using Proposition A.1, we obtain the following reformulation of the Lagrangian vorticity:

$$\Omega(\boldsymbol{a},t) = [J_{ij}]^{\mathrm{T}}\boldsymbol{\omega}(\boldsymbol{x}(\boldsymbol{a},t),t) = J[J]^{-1}\boldsymbol{\omega}(\boldsymbol{x}(\boldsymbol{a},t),t). \tag{6.1c}$$

Further, using Eqs. (6.1) and (A.8), we easily find

$$\boldsymbol{V}^{\mathrm{T}}(\boldsymbol{a},t)d\boldsymbol{a} = \boldsymbol{u}^{\mathrm{T}}(\boldsymbol{x},t)d\boldsymbol{x}, \tag{6.2a}$$

$$\Omega^{\mathrm{T}}(\boldsymbol{a},t)d\boldsymbol{s}^{(\boldsymbol{a})} = \boldsymbol{\omega}^{\mathrm{T}}(\boldsymbol{x},t)d\boldsymbol{s}^{(\boldsymbol{x})}, \tag{6.2b}$$

$$\Omega^{\mathrm{T}}(\boldsymbol{a},t)\boldsymbol{V}(\boldsymbol{a},t)dV^{(\boldsymbol{a})} = \boldsymbol{\omega}^{\mathrm{T}}(\boldsymbol{x},t)\boldsymbol{u}(\boldsymbol{x},t)dV^{(\boldsymbol{x})}. \tag{6.2c}$$

That is, the image fields are defined such that the elementary circulation $\boldsymbol{u}^{\mathrm{T}}d\boldsymbol{x}$, vorticity flux $\boldsymbol{\omega}^{\mathrm{T}}d\boldsymbol{s}^{(\boldsymbol{x})}$ and helicity $\boldsymbol{\omega}^{\mathrm{T}}\boldsymbol{u}dV^{(\boldsymbol{x})}$, are identical in the physical and the label space. In view of Eqs. (6.1a) and (6.1b), CIs (5.19), can be written as

$$\nabla_{\boldsymbol{a}} \times \frac{\partial \boldsymbol{V}(\boldsymbol{a},t)}{\partial t} = 0 \iff \frac{\partial}{\partial t}\left(\nabla_{\boldsymbol{a}} \times \boldsymbol{V}(\boldsymbol{a},t)\right) = 0, \quad \boldsymbol{a} \in V^{(\boldsymbol{a})}, \tag{6.3}$$

or

The vorticity image $\Omega = \Omega(\boldsymbol{a}, t)$ is time-independent, that is, $\Omega = \Omega(\boldsymbol{a})$. (6.4)

Further, Eq. (6.3) implies that the vector $\dot{\boldsymbol{V}}(\boldsymbol{a}, t)$ is curl-free, which leads to the representation

$$\dot{\boldsymbol{V}}(\boldsymbol{a}, t) = \nabla_{\boldsymbol{a}} \Phi^*(\boldsymbol{a}, t), \tag{6.5}$$

where $\Phi^*(\boldsymbol{a}, t)$ is a potential field.

*6.2. Further consequences of Cauchy's invariants*

In this Section we show how various known vorticity theorems and conservation laws, in the usual Eulerian formalism, can be derived by CIs. It is worth noticing that the derivations are straightforward and simple, rendering CIs a central result in vortical flows. Recall that the fluid is assumed *inviscid* and *barotropic*, *with conservative external forces.*

*D'Alembert-Euler Condition*: Employing Eqs. (A.2b) and (A.5), the following equation is easily proved

$$\frac{\partial \boldsymbol{V}(\boldsymbol{a}, t)}{\partial t} = [J]^{\mathrm{T}} \frac{d\boldsymbol{u}(\boldsymbol{x}, t)}{dt} + \nabla_{\boldsymbol{a}} \left( \frac{\dot{\boldsymbol{x}}^2(\boldsymbol{a}, t)}{2} \right), \tag{6.6}$$

where $d/dt$ is the material derivative applying to Eulerian fields. But since the vectors $\partial \boldsymbol{V}/\partial t$ and $d\boldsymbol{u}/dt$ are related as above, their corresponding rots will be related in accordance with Eq. (A.7), that is,

$$\nabla_{\boldsymbol{a}} \times \frac{\partial \boldsymbol{V}(\boldsymbol{a}, t)}{\partial t} = [J_{ij}]^{\mathrm{T}} \left( \nabla_{\boldsymbol{x}} \times \frac{d\boldsymbol{u}(\boldsymbol{x}, t)}{dt} \right) \tag{6.7}$$

Invoking then CIs (6.3), the left-hand side of Eq. (6.7) vanishes and, thus, we conclude that the fluid acceleration $d\boldsymbol{u}/dt$ is curl-free. This is known as the *D'Alembert-Euler condition*, and it is equivalent with the existence of an acceleration potential $\varphi^*(\boldsymbol{x}, t)$, such that [42] (Sec.16)

$$\frac{d\boldsymbol{u}(\boldsymbol{x}, t)}{dt} = \nabla_{\boldsymbol{x}} \varphi^*(\boldsymbol{x}, t). \tag{6.8}$$

*Beltrami's Vorticity Equation*: Invoking CIs, Eq. (6.4), Eq. (6.1c) is rewritten in the form

$$\Omega(\boldsymbol{a}) = J[J]^{-1}\boldsymbol{\omega}(\boldsymbol{x}, t). \tag{6.9}$$

Eq. (6.9) is known as the Cauchy vorticity formula [42] (Sec. 94). Since the left-hand side of Eq. (6.9) is time-independent, by differentiating with respect to time, invoking conservation of mass (2.9) and doing some simple manipulations, we obtain

$$[J]^{-1}\frac{d}{dt}\left(\frac{\boldsymbol{\omega}(\boldsymbol{x}, t)}{\rho(\boldsymbol{x}, t)}\right) + \frac{d[J]^{-1}}{dt}\frac{\boldsymbol{\omega}(\boldsymbol{x}, t)}{\rho(\boldsymbol{x}, t)} = 0. \tag{6.10}$$

In view of Eq. (A.4), the above reduces to

$$\frac{d}{dt}\left(\frac{\boldsymbol{\omega}(\boldsymbol{x}, t)}{\rho(\boldsymbol{x}, t)}\right) = \left(\frac{\boldsymbol{\omega}(\boldsymbol{x}, t)}{\rho(\boldsymbol{x}, t)} \cdot \nabla_{\boldsymbol{x}}\right)\boldsymbol{u}(\boldsymbol{x}, t). \quad \text{(in vector form)} \tag{6.11}$$

Eq. (6.11) is *Beltrami's vorticity equation* and expresses the evolution of vorticity in the physical domain [42] (Sec. 84).

*Ertel's Potential Vorticity Theorem*: Consider now any scalar quantity $S$ which is materially conserved, that is, $S \equiv S(\boldsymbol{a})$. Then, by CIs, Eq. (6.4), the product $\Omega^{\mathrm{T}}\nabla_{\boldsymbol{a}}S$ is time-independent, which, in conjunction with Eq. (6.1c), results in

$$\frac{d}{dt}\left[\left(J[J]^{-1}\boldsymbol{\omega}(\boldsymbol{x}, t)\right)^{\mathrm{T}}\nabla_{\boldsymbol{a}}S(\boldsymbol{a})\right] = 0. \tag{6.12}$$

Now, using the transformation rule for the gradient operator, Eq. (A.2a), invoking conservation of mass (2.9), and making some easy manipulations, we get

$$\frac{d}{dt}\left(\frac{\boldsymbol{\omega}(\boldsymbol{x}, t)}{\rho(\boldsymbol{x}, t)} \cdot \nabla_{\boldsymbol{x}}S(\boldsymbol{x}, t)\right) = 0, \quad \text{(in vector form)} \tag{6.13}$$

which is *Ertel's Potential Vorticity Theorem* [42] (Sec. 102).

*Hankel-Kelvin Circulation Theorem*: Consider an arbitrary closed material loop in the label space, denoted by $C^{(\boldsymbol{a})}$ (time-independent by definition). In view of Eq. (6.2a), the circulation $\Gamma \equiv \Gamma(t)$ is given by

$$\Gamma(t) = \oint_{C^{(\boldsymbol{x})}} \boldsymbol{u}^{\mathrm{T}}(\boldsymbol{x}, t)d\boldsymbol{x} = \oint_{C^{(\boldsymbol{a})}} \boldsymbol{V}^{\mathrm{T}}(\boldsymbol{a}, t)d\boldsymbol{a}, \tag{6.14}$$

where $C^{(\boldsymbol{x})}$ is the image of the contour $C^{(\boldsymbol{a})}$ in the physical space. Invoking Stokes' theorem and using Eq. (6.1b) and CIs in the form of Eq. (6.4), we obtain

$$\oint_{C^{(\boldsymbol{a})}} \boldsymbol{V}^{\mathrm{T}}(\boldsymbol{a},t)d\boldsymbol{a} = \int_{S^{(\boldsymbol{a})}} (\nabla_{\boldsymbol{a}} \times \boldsymbol{V}(\boldsymbol{a},t))^{\mathrm{T}} d\boldsymbol{s}^{(\boldsymbol{a})} = \int_{S^{(\boldsymbol{a})}} \Omega^{\mathrm{T}}(\boldsymbol{a})d\boldsymbol{s}^{(\boldsymbol{a})}, \quad (6.15)$$

where $S^{(\boldsymbol{a})}$ is a surface with boundary $C^{(\boldsymbol{a})}$. Eq. (6.15), in conjunction with Eq. (6.14), results in $d\Gamma(t)/dt = 0$, which is the *Hankel-Kelvin Circulation Theorem* [42] (Sec. 49).

*Moffatt's Helicity Conservation Theorem*: By definition, the helicity over a fluid region is

$$H(t) = \int_{V^{(\boldsymbol{x})}} \boldsymbol{\omega}^{\mathrm{T}}(\boldsymbol{x},t)\boldsymbol{u}(\boldsymbol{x},t)dV^{(\boldsymbol{x})} = \int_{V^{(\boldsymbol{a})}} \Omega^{\mathrm{T}}(\boldsymbol{a})\boldsymbol{V}(\boldsymbol{a},t)dV^{(\boldsymbol{a})}, \quad (6.16)$$

where the second equality is obtained using Eq. (6.2c). Differentiating in time the above equation, and employing CIs, Eqs. (6.4) and (6.5), we obtain

$$\frac{dH(t)}{dt} = \int_{V^{(\boldsymbol{a})}} \Omega^{\mathrm{T}}(\boldsymbol{a})\dot{\boldsymbol{V}}(\boldsymbol{a},t)dV^{(\boldsymbol{a})} = \int_{V^{(\boldsymbol{a})}} \Omega^{\mathrm{T}}(\boldsymbol{a})\nabla_{\boldsymbol{a}}\Phi^{*}(\boldsymbol{a},t)dV^{(\boldsymbol{a})}. \quad (6.17)$$

Since, by construction, the Lagrangian vorticity is divergence-free, $\nabla_{\boldsymbol{a}}\Omega^{\mathrm{T}} = 0$, an application of the divergence theorem to the last term of the above equation leads to

$$\frac{dH(t)}{dt} = \int_{\partial V^{(\boldsymbol{a})}} \Phi^{*}(\boldsymbol{a},t)\Omega^{\mathrm{T}}(\boldsymbol{a})d\boldsymbol{s}^{(\boldsymbol{a})}. \quad (6.18)$$

Now, invoking Eq. (6.2b), we see that

$$\Omega^{\mathrm{T}}(\boldsymbol{a})d\boldsymbol{s}^{(\boldsymbol{a})} = 0 \text{ on } \partial V^{(\boldsymbol{a})} \iff \boldsymbol{\omega}^{\mathrm{T}}(\boldsymbol{x},t)d\boldsymbol{s}^{(\boldsymbol{x})} = 0 \text{ on } \partial V^{(\boldsymbol{x})}, \quad (6.19)$$

which, in conjunction with Eq. (6.18), leads to the conclusion that the helicity is conserved, $dH(t)/dt = 0$, over any region of the fluid (either in the physical space or in the label space) such that the vorticity is everywhere tangent to its boundary. This is *Moffatt's Helicity Conservation Theorem*, proved by Moffatt [39] (p. 112).

## Appendix: On the Jacobian matrix - Derivatives and related formulae

The Jacobian matrix of transformation (2.1), denoted by $[J]$, is the fundamental entity in the Lagrangian description of the fluid flow. In this paper, we adopt the following definition:

$$[J] \equiv \frac{D(x_1,x_2,x_3)}{D(a_1,a_2,a_3)} \equiv \begin{pmatrix} \partial x_1/\partial a_1 & \partial x_1/\partial a_2 & \partial x_1/\partial a_3 \\ \partial x_2/\partial a_1 & \partial x_2/\partial a_2 & \partial x_2/\partial a_3 \\ \partial x_3/\partial a_1 & \partial x_3/\partial a_2 & \partial x_3/\partial a_3 \end{pmatrix} =$$

$$= \left( \begin{pmatrix} \partial/\partial a_1 \\ \partial/\partial a_2 \\ \partial/\partial a_3 \end{pmatrix} (x_1 x_2 x_3) \right)^{\mathrm{T}} \equiv \left(\nabla_{\boldsymbol{a}} \boldsymbol{x}^{\mathrm{T}}\right)^{\mathrm{T}}.$$

By transposition we get $[J]^{\mathrm{T}} = \begin{pmatrix} \partial/\partial a_1 \\ \partial/\partial a_2 \\ \partial/\partial a_3 \end{pmatrix} (x_1 x_2 x_3) \equiv \nabla_{\boldsymbol{a}} \boldsymbol{x}^{\mathrm{T}}$, and by using chain rule we find the following important formula:

$$\nabla_{\boldsymbol{a}} (\cdot) = [J]^{\mathrm{T}} \nabla_{\boldsymbol{x}} (\cdot)\,, \tag{A.1}$$

which relates the gradient operators $\nabla_{\boldsymbol{a}}$ and $\nabla_{\boldsymbol{x}}$. Applying Eq. (A.1) to a scalar field $S$ and a vector field $\boldsymbol{b}$, respectively, we obtain

$$\nabla_{\boldsymbol{a}} S(\boldsymbol{a},t) = [J]^{\mathrm{T}} \nabla_{\boldsymbol{x}} S(\boldsymbol{x},t), \tag{A.2a}$$

$$\nabla_{\boldsymbol{a}} \boldsymbol{b}^{\mathrm{T}}(\boldsymbol{a},t) = [J]^{\mathrm{T}} \nabla_{\boldsymbol{x}} \boldsymbol{b}^{\mathrm{T}}(\boldsymbol{x},t). \tag{A.2b}$$

Taking the time derivative of each element of the Jacobian matrix, and exploiting Eq. (A.1), we find

$$\frac{d[J]^{\mathrm{T}}}{dt} = \nabla_{\boldsymbol{a}} \dot{\boldsymbol{x}}^{\mathrm{T}} = [J]^{\mathrm{T}} \nabla_{\boldsymbol{x}} \boldsymbol{u}^{\mathrm{T}}, \quad \frac{d[J]}{dt} = \left(\nabla_{\boldsymbol{x}} \boldsymbol{u}^{\mathrm{T}}\right)^{\mathrm{T}} [J]. \tag{A.3}$$

By using the identity $d[J]^{-1}/dt = -[J]^{-1}\,(d[J]/dt)\,[J]^{-1}$, and Eqs. (A.3), we get the following useful relation

$$\frac{d[J]^{-1}}{dt} = -[J]^{-1} \left(\nabla_{\boldsymbol{x}} \boldsymbol{u}^{\mathrm{T}}\right)^{\mathrm{T}}. \tag{A.4}$$

Further, by direct computations we obtain the formula

$$\left(\nabla_{\boldsymbol{a}} \dot{\boldsymbol{x}}^{\mathrm{T}}\right) \dot{\boldsymbol{x}} = \frac{1}{2} \nabla_{\boldsymbol{a}} \left(\dot{\boldsymbol{x}}^{\mathrm{T}} \dot{\boldsymbol{x}}\right). \tag{A.5}$$

Another interesting result, which is used in Sec. 6, is given below:
**Proposition A.1.** Assume that the point transformation (2.1) is a $C^2$ bi-diffeomorphism, and that the vector fields $\boldsymbol{Q} = \boldsymbol{Q}(\boldsymbol{a}, t)$, $\boldsymbol{q} = \boldsymbol{q}(\boldsymbol{x}, t)$ and the scalar field $F = F(\boldsymbol{a}, t)$ are $C^1$. Then, if the fields $\boldsymbol{Q}$, $\boldsymbol{q}$ and $F$ are related by the equation

$$\boldsymbol{Q}(\boldsymbol{a}, t) = [J]^{\mathrm{T}}\boldsymbol{q}(\boldsymbol{x}, t) + \nabla_{\boldsymbol{a}} F(\boldsymbol{a}, t), \tag{A.6}$$

the vector fields $\nabla_{\boldsymbol{a}} \times \boldsymbol{Q}$ and $\nabla_{\boldsymbol{x}} \times \boldsymbol{q}$ are related as follows

$$\nabla_{\boldsymbol{a}} \times \boldsymbol{Q}(\boldsymbol{a}, t) = [J_{ij}]^{\mathrm{T}} \left(\nabla_{\boldsymbol{x}} \times \boldsymbol{q}(\boldsymbol{x}, t)\right). \ \blacksquare \tag{A.7}$$

Variants of Proposition A.1 can be found in [40, 41], [39](Appendix A4).

*Transformation of geometric elements*: The point transformation (2.1) induces transformations between the line, surface and volume elements in the label space and in the physical space. The corresponding formulae are given below [42] (Sec.15)

$$d\boldsymbol{x} = [J] d\boldsymbol{a}, \tag{A.8a}$$

$$d\boldsymbol{s}^{(\boldsymbol{x})} = [J_{ij}] d\boldsymbol{s}^{(\boldsymbol{a})}, \tag{A.8b}$$

$$dV^{(\boldsymbol{x})} = J dV^{(\boldsymbol{a})}. \tag{A.8c}$$